\documentclass[graybox]{svmult}

\usepackage{mathptmx}       
\usepackage{helvet}         
\usepackage{courier}        
\usepackage{type1cm}        
\usepackage{makeidx}         
\usepackage{multicol}        
\usepackage[bottom]{footmisc}

\usepackage{natbib}
\bibpunct{(}{)}{;}{a}{}{,}

\makeindex             

\usepackage[english]{babel}
\usepackage{amsmath}
\usepackage{amsfonts}
\usepackage{amssymb}
\usepackage{graphicx,subfigure}
\usepackage{listings}

\usepackage[colorlinks=true,citecolor=blue,urlcolor=green]{hyperref}

\definecolor{codegreen}{rgb}{0,0.6,0}
\definecolor{codegray}{rgb}{0.5,0.5,0.5}
\definecolor{codepurple}{rgb}{0.58,0,0.82}
\definecolor{backcolour}{rgb}{0.95,0.95,0.92}
 
\lstdefinestyle{mystyle}{
	moredelim=[is][\bfseries]{[*}{*]},
    backgroundcolor=\color{backcolour},   
    commentstyle=\color{codegreen},
    keywordstyle=\color{magenta},
    numberstyle=\tiny\color{codegray},
    stringstyle=\color{codepurple},
    basicstyle=\footnotesize,
    breakatwhitespace=false,         
    breaklines=true,                 
    captionpos=b,                    
    keepspaces=true,                 
    numbers=left,                    
    numbersep=5pt,                  
    showspaces=false,                
    showstringspaces=false,
    showtabs=false,                  
    tabsize=2
} 

\makeatletter
\newcommand*\mysize{%
  \@setfontsize\mysize{4.8}{8}%
}
\makeatother

\makeatletter
\newcommand\footnoteref[1]{\protected@xdef\@thefnmark{\ref{#1}}\@footnotemark}
\makeatother

\begin{document}

\title*{Asteroseismic Stellar Modelling with AIMS}
\author{Mikkel N.~Lund and Daniel R.~Reese}

\institute{Mikkel N.~Lund \at School of Physics and Astronomy, University of Birmingham, Birmingham B15 2TT, UK, \\
\email{lundm@bison.ph.bham.ac.uk}
\and Daniel R.~Reese \at LESIA, Observatoire de Paris, PSL Research University, CNRS, Sorbonne Universit\'{e}s, UPMC Univ.~Paris 06, Univ.~Paris Diderot, Sorbonne Paris Cit\'{e}e, 92195 Meudon, France, \\ \email{daniel.reese@obspm.fr}}

%
\maketitle

\vspace{-3cm}

\abstract{The goal of \textsc{aims} (Asteroseismic Inference on a Massive Scale) is to estimate stellar parameters and credible intervals/error bars in a Bayesian manner from a set of asteroseismic
frequency data and so-called classical constraints. To achieve reliable parameter
estimates and computational efficiency, it searches through a grid of
pre-computed models using an MCMC algorithm --- interpolation within the grid of
models is performed by first tessellating the grid using a Delaunay
triangulation and then doing a linear barycentric interpolation on matching
simplexes. Inputs for the modelling consist of individual frequencies from
peak-bagging, which can be complemented with classical spectroscopic constraints.
\textsc{aims} is mostly written in {\ttfamily Python} with a modular structure to facilitate
contributions from the community.  Only a few computationally intensive parts
have been rewritten in {\ttfamily Fortran} in order to speed up calculations.}

\section{Introduction}
\label{sec:intro}

The \textsc{aims} (Asteroseismic Inference on a Massive Scale) software was developed by D.~R.~Reese as one of the deliverables for the SPACEINN network, a European project specialised in helio-
and asteroseismology. The goal of this software is to estimate stellar
parameters and reliable error bars for a given set of asteroseismic and classical
constraints.  The present tutorial explains how to use this software through
various simple examples.  Specifically, it explains how to find stellar
parameters and error bars for a given set of constraints, generate a binary grid
file usable by \textsc{aims}, and test the accuracy of the interpolation for a
given grid. The necessary files and data to run the examples in this tutorial
are available at the following website:
\url{http://bison.ph.bham.ac.uk/spaceinn/aims/tutorial/}.

\section{Getting started}
\label{sec:pre}

\subsection{Prerequisites and downloads for the tutorial}

The following will be needed to run the examples in this tutorial:
\begin{itemize}
\item \textbf{Python modules.}  The following {\ttfamily Python} modules and utilities are
      needed by \textsc{aims}: {\ttfamily emcee}, {\ttfamily corner}, {\ttfamily dill}, 
      {\ttfamily Scipy}, {\ttfamily Numpy}, {\ttfamily f2py},
      {\ttfamily Matplotlib}.  The last four are included in most distributions.
\item \textbf{A grid of models.} \textsc{aims} works by comparing observational
      data to a grid of pre-computed models. The tutorial website provides two
      binary grids ({\ttfamily data\_mesa} and {\ttfamily data\_cestam}), which will
      be used for finding model fits to a set of observed stars, as well as a
      folder containing a non-binary subset of one of the grids, which we shall
      use when trying to generate a binary grid. When ``unpacked'' by
      \textsc{aims}, some of the grids take up a lot of live memory\footnote{In some cases, this problem can further be compounded by
      the use of parallelisation, which is activated by setting
      \texttt{parallel=True} in \texttt{AIMS$\_$configure.py}.}
      (RAM).  Accordingly, a
      ``light'' version of the \textsc{cestam} grid,  \texttt{data$\_$cestam$\_$reduced},
      has been provided.
\item \textbf{File(s) with observational data.} The tutorial website provides
      files with observational data for three stars (Stars 1--3). The mode
      frequencies were obtained from peak-bagging of \textit{Kepler} data for
      the so-called LEGACY project \citep{2016arXiv161200436L}. Spectroscopic
      data were obtained from the Stellar Parameters Classification tool
      \citep[\textsc{spc};][]{2012Natur.486..375B}.
\end{itemize}

\subsection{Downloading and installing AIMS}
The latest version of the \textsc{aims} package, currently version 1.2, can
be downloaded from the following site:
\url{http://bison.ph.bham.ac.uk/spaceinn/aims/version1.2/index.html}. 
This file is unpacked as follows:
\begin{lstlisting}
tar -zxvf AIMS.tgz
\end{lstlisting}
The \textsc{aims} program itself is contained within the {\ttfamily AIMS} folder.

As mentioned earlier, the latest version of \textsc{aims} contains some {\ttfamily Fortran}
subroutines which need to be compiled before running \textsc{aims}.  This is
done via the {\ttfamily f2py} program.  A {\ttfamily Makefile} has been provided for convenience. Please
edit the {\ttfamily Makefile} by inserting the appropriate {\ttfamily Fortran} compiler and compilation
options.  Then run the following command:
\begin{lstlisting}
make
\end{lstlisting}
This will produce a file called \texttt{aims$\_$fortran.so} which can be used by
\textsc{aims}.

\section{Running AIMS: model fit}
\label{sec:run}
We shall first consider running \textsc{aims} with the goal of optimising the
fit to a given set of observational constraints, such as mode frequencies,
ratios, and spectroscopic parameters. In Sect.~\ref{sec:int} we will look into
testing the interpolation scheme in \textsc{aims}. Figure 1.1 of the
overview document\footnote{\label{note1}\url{http://bison.ph.bham.ac.uk/spaceinn/aims/version1.2/_downloads/Overview.pdf}} provides a simple schematic flowchart with the basic working
components of \textsc{aims}.


\subsection{Setting up the configuration file}
Before running the \textsc{aims} program, a binary file with the grid must first
be created --- we will come to this in Sect.~\ref{sec:bin}. Assuming in the mean
time that has been done, the most important concern is to set up the
configuration file: \texttt{AIMS$\_$configure.py}.

Most parameters in the configuration file are well documented and should be
self-explanatory --- for instance, you can choose which asteroseismic parameters to 
fit\footnote{The user should be careful not to choose a set of asteroseismic parameters
which are redundant, as this would lead to a singular covariance matrix and poor
numerical results.} (individual frequencies, ratios, or average asteroseismic
parameters), the name of the binary grid to use, which parameters should be
output from the optimisation, control parameters for the MCMC, which grid
parameters to use in the optimisation and which priors to set on these etc.

Two parameters are of special relevance:
\begin{itemize}
\item \texttt{write$\_$data} should be set to \texttt{False} (see
      Sect.~\ref{sec:bin} for when this should be \texttt{True}).
\item \texttt{test$\_$interpolation} should be set to \texttt{False} (see
      Sect.~\ref{sec:int} for when this should be \texttt{True}).
\end{itemize}

It is also very important to put the correct values for the
\texttt{grid$\_$params} and \texttt{user$\_$params} parameters.  These values
will depend on the binary grid being used.  They are provided on the tutorial
website and can also be obtained with the {\ttfamily analyse$\_$grid.py} utility\footnote{\url{http://bison.ph.bham.ac.uk/spaceinn/aims/tutorial/download/analyse_grid.py}}, which should be run in the {\ttfamily AIMS} folder.

\subsection{Observational constraints}
The file with observational constraints follows a very similar format to the one
used with the Asteroseismic Modeling Portal\footnote{See \url{https://amp.phys.au.dk/guide/fileformat}.} (AMP), apart from a few minor differences\footnote{See
\url{http://bison.ph.bham.ac.uk/spaceinn/aims/version1.2/formats.html\#format-of-a-file-with-observational-constraints}.}. This format is described as follows:

\begin{itemize}
\item Characters following a ``{\ttfamily \#}'' are ignored, and the ordering of the lines is
      unimportant.
\item A set of lines, one per individual mode, describe the asteroseismic observables.
      If the keyword \texttt{read$\_$n} is set to \texttt{True}, then four
      columns should be given, namely, the degree of the mode ($l$), the radial
      order ($n$), the frequency in cyclic $\mu$Hz ($f$), and its associated
      error bar ($\delta f$). If \texttt{read$\_$n} is \texttt{False}, then only
      $l$, $f$, and $\delta f$ should be given. Note that in \textsc{aims}, even
      if you choose to work with frequency combinations such as ratios, the
      asteroseismic inputs are still individual frequencies --- \textsc{aims} will use
      these to calculate the frequency combinations, associated error bars, and
      correlations.
\item Optionally, non-asteroseismic constraints may be included in addition to the
      asteroseismic observables. The first column must consist of a character or a
      keyword, e.g., ``{\ttfamily T} or {\ttfamily Teff}" ($T_{\rm eff}$), ``{\ttfamily L} or {\ttfamily Luminosity}" ($L/{\rm L}_\odot$), ``{\ttfamily R} or
      {\ttfamily Radius}" ($R/{\rm R}_\odot$), ``{\ttfamily M}, {\ttfamily Fe$\_$H} or {\ttfamily M$\_$H}" ([M/H]), ``{\ttfamily g} or {\ttfamily log$\_$g}" ($\log
      g$), or ``{\ttfamily Rho}'' ($\rho$). This is followed by a central value and an error.
\item If a non-asteroseismic constraint is included, e.g., ``{\ttfamily Teff  5777  50}'', then
      it is assumed that this parameter follows a Gaussian distribution.
      Alternatively, you may explicitly define the distribution to be used,
      e.g., ``{\ttfamily Teff  Uniform  5727  5837}'' meaning a uniform distribution
      from 5727 to 5837 K. You can adopt either a ``{\ttfamily Uniform}", ``{\ttfamily Gaussian}" (the
      default), or ``{\ttfamily Truncated$\_$gaussian}'' probability distribution. 
\item The average asteroseismic parameters $\nu_{\rm max}$ (``\texttt{numax}'') and
      $\Delta\nu$ (``\texttt{Dnu}'') can also be supplied here as a
      constraint\footnote{We note that this is not the preferred way of
      supplying $\Delta\nu$, as it does not correctly take into account
      correlations with other asteroseismic constraints. A better approach is to
      introduce the large separation via the \texttt{seismic$\_$constraints}
      variable in \texttt{AIMS$\_$configure.py}.}.
\end{itemize}

The following gives an example of how the constraints file may look like:
\begin{lstlisting}
0 1847.63576435 0.76075810437 
1 1904.67149676 0.827874359443 
2 1954.85689326 0.751669075869 
0 1968.18957034 0.212209047899 
           ...
T 6120 80
Fe_H -0.06 0.12
numax 2763 100
\end{lstlisting}

\subsection{Running AIMS}

Once the grid and configuration file are set up, one simply runs the program as
follows:
\begin{lstlisting}
./AIMS.py observational_constraints_file
\end{lstlisting}
\textsc{aims} will then import the \texttt{Python} configuration file,
\texttt{AIMS$\_$configure.py}, so make sure you do not modify the name of this
file.  Output generated from \textsc{aims} will then be saved to a run folder with
the same name as your constraints file, so give your constraints file a sensible
name so that you can distinguish the results from several runs for the same
star.  The variable \texttt{output$\_$dir} in \texttt{AIMS$\_$configure.py}
specifies the path to the root folder which contains all of the run folders.

\subsection{Understanding the results}

\begin{figure*}[t]
    \centering
    \begin{subfigure}
        \centering
        \includegraphics[width=0.49\textwidth]{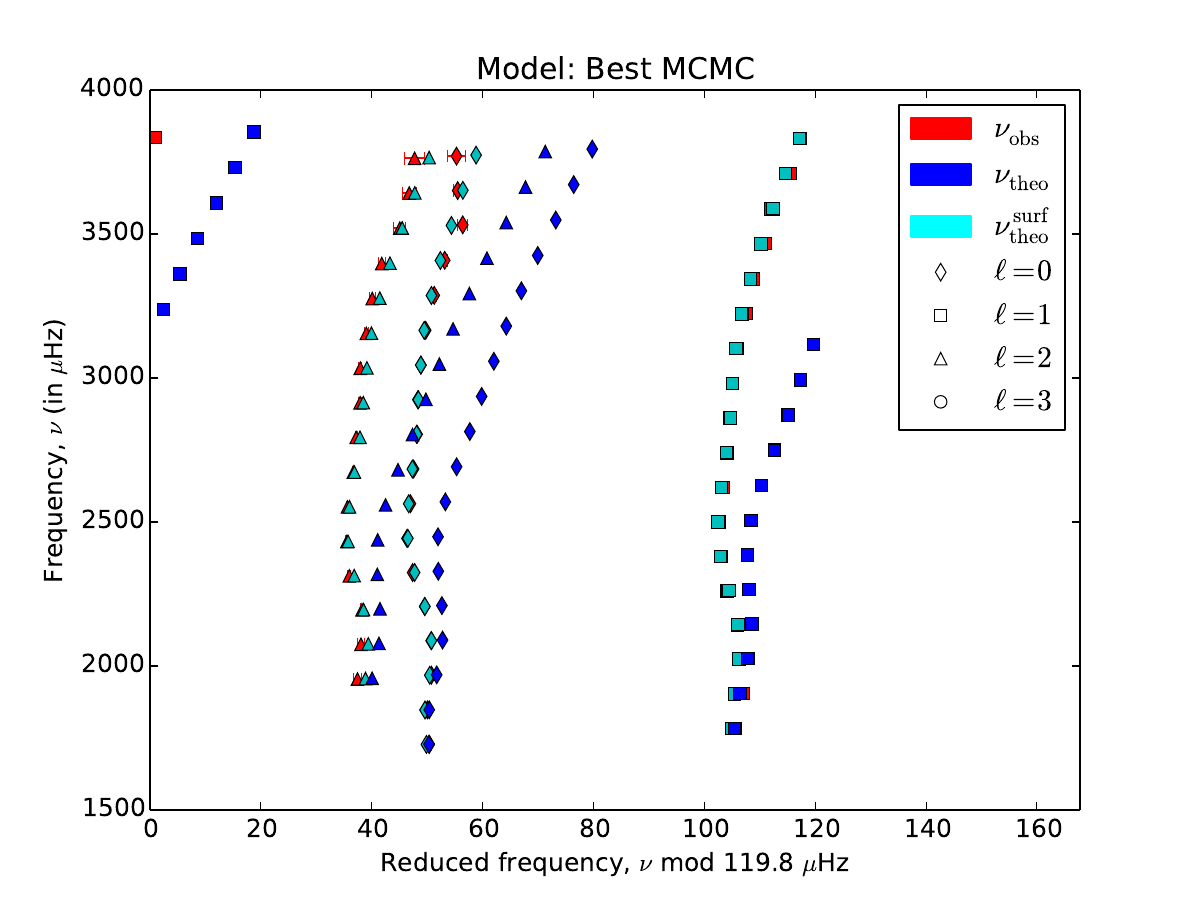}
    \end{subfigure}\hfill
    \begin{subfigure}
        \centering
        \includegraphics[width=0.49\textwidth]{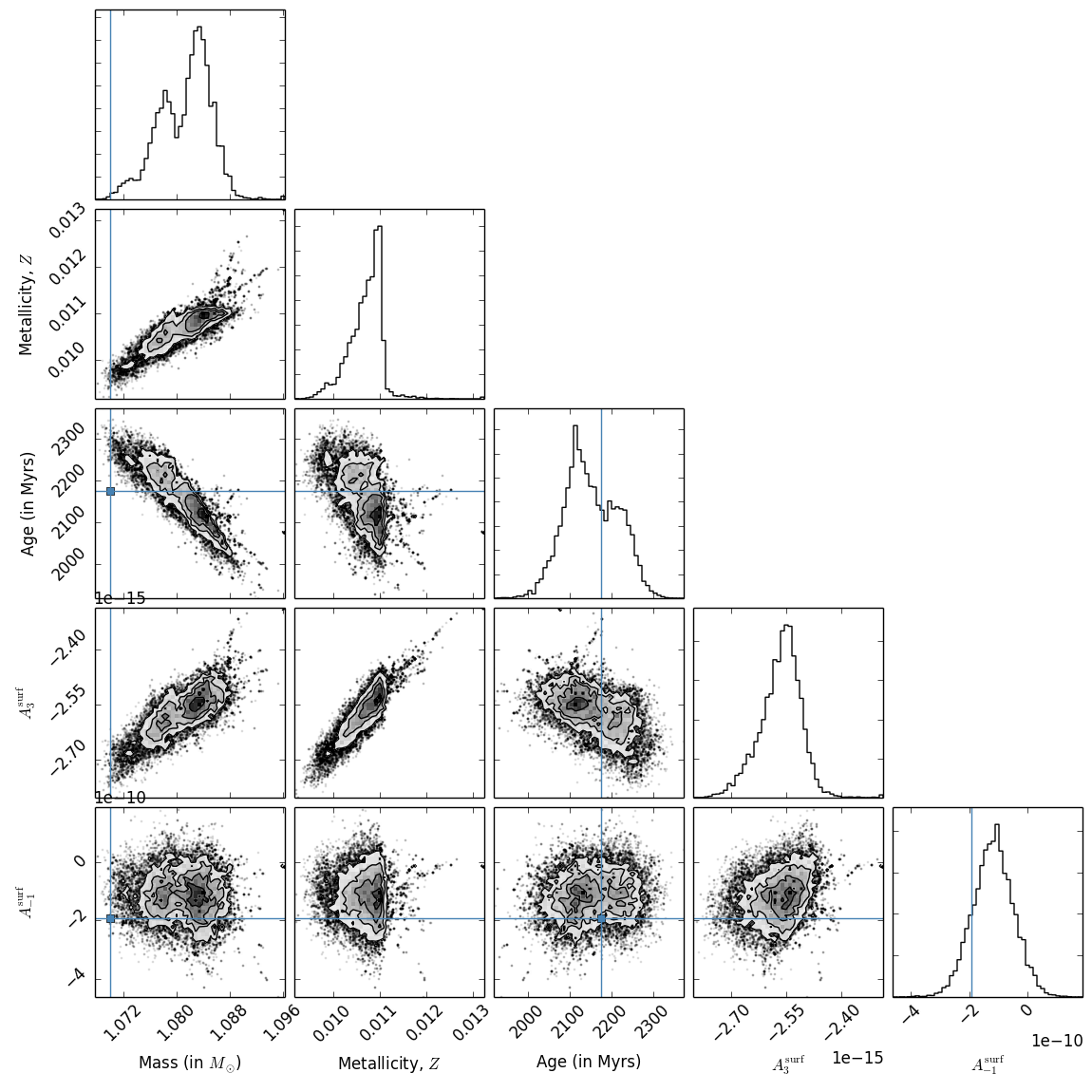}
    \end{subfigure}
    \caption{Example of output plots from an \textsc{aims} run.
    \textit{Left:} \'{E}chelle diagram for the best MCMC model showing the
    observed frequencies, theoretical frequencies, and surface-corrected
    theoretical frequencies. \textit{Right:} Triangle plot with correlation maps
    between different stellar parameters for the MCMC samples. The blue
    lines indicate the results obtained from the initial full-grid search.}
\label{fig:fig1}
\end{figure*}

The results from \textsc{aims} are obtained from the posterior distributions of
the MCMC run on the model grid, as illustrated in the right panel of
Fig.~\ref{fig:fig1}. In \texttt{AIMS$\_$configure.py}, you can define the set
of stellar parameters for which you want an output, as well as which plots to
create.  For computational reasons, such parameters are only calculated for a
subset of the MCMC samples, except for those parameters actively used in the
MCMC optimisation.  Accordingly, \textsc{aims} produces two files with the
samples: \texttt{samples.txt} with all of the MCMC samples but only the
parameters involved in the optimisation, and \texttt{samples$\_$big.txt} with a
subset of the samples but all of the stellar parameters.  The corresponding
files \texttt{results.txt} and \texttt{results$\_$big.txt} provide summary
statistics for the above samples, namely, the distribution averages and standard
deviations, along with correlations between different model parameters.  The
samples files may, of course, be used to extract different summary statistics
(e.g., median or mode) for the parameters. The file
\texttt{best$\_$MCMC$\_$model.txt} gives the stellar parameters and computed
mode frequencies for the best model from the MCMC, whereas
\texttt{best$\_$grid$\_$model.txt} gives the best model from the initial grid
search used to initialise the MCMC (i.e., prior to interpolation within
the grid).  The left panel of Fig.~\ref{fig:fig1} shows the {\'e}chelle diagram
produced from the observations and from the best MCMC model.

\section{Creating your own binary grid files with AIMS}
\label{sec:bin}

To compute a binary grid file, you first need to calculate a grid of models with
your favourite stellar evolution code, as well as oscillation modes for each
model. Information for each model should then be entered into a ``model list"
file. The first line of this file should contain a prefix which is typically
the root folder of the grid of models and, optionally, a postfix giving the
end part of the filenames with the mode frequencies, the default value being
``{\ttfamily .freq}''. The following lines then contain multiple columns with the
information for each model in the grid. Below we show an example of a model list file (see column description in Table \ref{tab:columns_model_list}):
\begin{lstlisting}
/home/dreese/models_inversions/Grid_6819/m1.6.ovh0.0.ovhe0.00.z0.01756.y0.26/
m1.6.z0.01756.y0.26_n2026.FGONG  3.18272E+33 9.411631E+11 2.689504E+35 0.01756 0.72244 2.402848E+03 4543.38696
m1.6.z0.01756.y0.26_n2093.FGONG  3.18272E+33 9.645173E+11 2.811105E+35 0.01756 0.72244 2.402920E+03 4537.93601
m1.6.z0.01756.y0.26_n1986.FGONG  3.18272E+33 9.010663E+11 2.513558E+35 0.01756 0.72244 2.402825E+03 4565.49605
m1.6.z0.01756.y0.26_n1575.FGONG  3.18272E+33 1.166064E+12 3.745062E+35 0.01756 0.72244 2.402670E+03 4434.01896
\end{lstlisting}

\begin{table}[t]
\begin{center}
\caption{Columns in ``model list'' file}
\label{tab:columns_model_list}
\resizebox{\textwidth}{!}{\mysize\begin{tabular}{lccccccccc}
\hline
\textbf{Column \#}  &  1 & 2 & 3 & 4 & 5 & 6 & 7 & 8 & 9, 10 \ldots \\
\hline
\textbf{Parameter} & Model name & Mass & Rad. & Lum. & $Z_0$ & $X_0$ & Age & $T_{\mathrm{eff}}$ & \texttt{user$\_$params} \\
\textbf{Unit}          &  \ldots   & g    & cm     & g cm$^2$s$^{-3}$ & \ldots & \ldots & Myr & K & \ldots \\
\hline
\end{tabular}}
\end{center}
\end{table}

The prefix plus each model name in the first column, plus the postfix gives the
name of a file that contains the oscillation parameters of the model.  These
files can come in one of two formats, as specified by the
\texttt{mode$\_$format} variable in \texttt{AIMS$\_$configure.py}.  One of the
formats is a {\ttfamily Fortran} binary format known as the ``grand summary'' file from the \textsc{adipls}\footnote{\url{http://astro.phys.au.dk/~jcd/adipack.n/}} code
\citep{Christensen-Dalsgaard2008} and is described on pages 32--33 of the \textsc{adipls} documentation\footnote{\url{http://astro.phys.au.dk/~jcd/adipack.n/notes/adiab_prog.ps.gz}}.
The other format is a text format (described below) and which is what is
used in this tutorial.

The text version of the oscillation parameter files begins with a one-line
header followed by five columns which correspond to $l$, $n$, frequency,
{\ttfamily dfreq$\_$var}, and mode inertia. Note that the {\ttfamily dfreq$\_$var} column is currently
discarded, as are frequencies above the estimated cut-off frequency times the
value of the \texttt{cutoff} variable in \texttt{AIMS$\_$configure.py}.

To generate a binary grid you should specify to following relevant parameters in
\texttt{AIMS$\_$configure.py}:
\begin{itemize}
\item \texttt{write$\_$data} should be set to \texttt{True}.
\item \texttt{list$\_$grid} gives the filename of the model list file (see
      above).
\item \texttt{binary$\_$grid} gives the filename of the binary file that will
      be generated. If \texttt{write$\_$data} is set to \texttt{False}, this is
      the binary grid that will be loaded.
\item \texttt{grid$\_$params} specifies the parameters relevant to the grid you
      want to generate (excluding age, which will be dealt with separately).
      \textit{It is extremely important that each set of values for these
      parameters corresponds to a unique evolutionary track, since}
      \textsc{aims} \textit{reconstructs these tracks based on these values.}
      For instance, if mass and $Z$ are the parameters which describe your grid,
      then pairs of values such as $(\mathrm{mass}=1\,\mathrm{M}_{\odot},
      Z=0.02)$ should correspond to a unique track.
\item \texttt{user$\_$params} specifies supplementary parameters which describe
      your grid.  This variable should contain a pair of strings for each
      supplementary parameter.  The first string should give the parameter's
      name\footnote{The names ``{\ttfamily Xs}'' and ``{\ttfamily Zs}'' should be used for the surface
      hydrogen and metallicity content, as some of the functions in
      \textsc{aims} specifically look for these variables.}.  The second string
      should be a nice \LaTeX version of the name to be used in plot titles.
\item \texttt{npositive} should be set to \texttt{True} if you only want to save
      $n\geq 0$ modes (i.e., acoustic modes) in the binary file.
\item \texttt{cutoff}: frequencies above this value times the estimated cut-off
      frequency (as based on a scaling law) will be discarded.  For example, if
      \texttt{cutoff=1.1} then only frequencies below
      $1.1\nu_{\rm{c}}$ are kept.
\item \texttt{agsm$\_$cutoff}: this only applies to binary ``grand summary''
      files from \textsc{adipls}.  If set to \texttt{True}, then only frequencies for
      which \texttt{icase=10010} (i.e., which are below the cut-off
      frequency when using an isothermal boundary condition) are kept.
\end{itemize}

The binary file is then simply generated by running:
\begin{lstlisting}
./AIMS.py
\end{lstlisting}

\section{Testing interpolation accuracy}
\label{sec:int}

\subsection{Calculating interpolation errors}
Since \textsc{aims} works by interpolating in a pre-computed grid of stellar
models, it also includes a way of testing the accuracy of the interpolation.
To test the interpolation you should specify to following options in
\texttt{AIMS$\_$configure.py}:
\begin{itemize}
\item \texttt{write$\_$data} should be set to \texttt{False}.
\item \texttt{test$\_$interpolation} should be set to \texttt{True}.
\item \texttt{interpolation$\_$file} gives the name of the binary file that
      will contain the output of the test. The results saved in this
      file can then be plotted using \texttt{plot$\_$interpolation$\_$test.py}. 
\end{itemize}

The interpolation test is then simply run as:
\begin{lstlisting}
./AIMS.py
\end{lstlisting}

\subsection{Various interpolation errors}

\begin{figure}[t]
    \centering
    \includegraphics[width=0.7\textwidth]{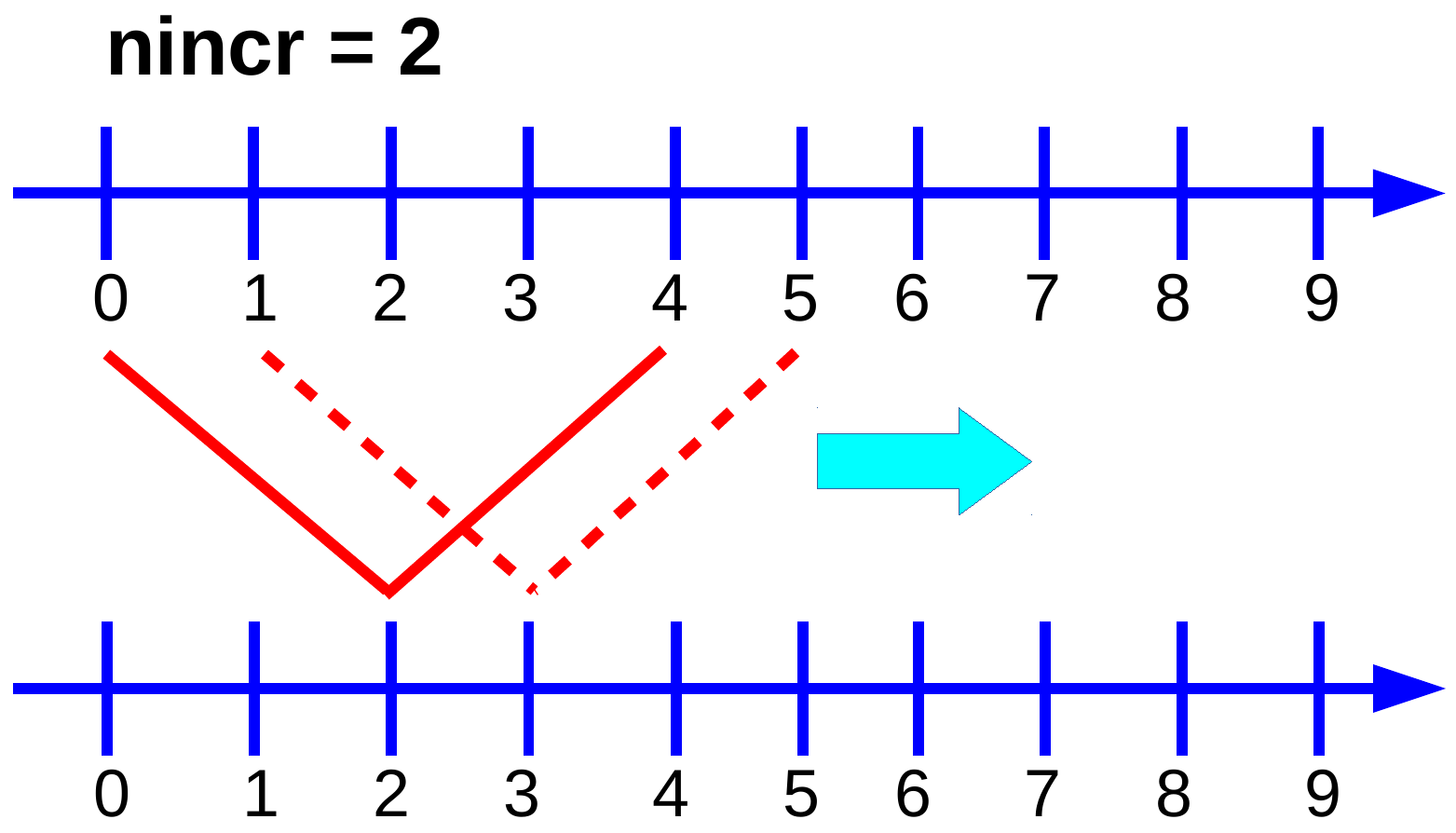}
    \caption{Schematic plot showing how the age interpolation
    tests are calculated for \texttt{nincr=2}. Successive models along the track are not
    necessarily equally spaced in age.\label{fig:age_interpolation}}
\end{figure}

There are two basic components to model interpolation in \textsc{aims}, each
of which contributes to interpolation error:
\begin{itemize}
\item \textbf{Age interpolation:} this is interpolation along a given
      evolutionary track.
\item \textbf{Track interpolation:} this is interpolation as a function of the
      other model parameters, such as mass, metallicity, mixing length
      parameter, or whatever parameters are relevant to your grid.
\end{itemize}
The first type of interpolation is dealt with through a simple linear
interpolation between two adjacent models on the evolutionary track. The second
uses Delaunay tessellation before calculating linear barycentric weights.
For a more detailed description of interpolation
in \textsc{aims}, we refer the reader to chapter 4 of the overview document\footnoteref{note1}.

The interpolation tests carried out in \textsc{aims} allow the user to estimate
the error from both types of interpolation.  For the age interpolation, we
number the models on a given evolutionary track, starting at $n=0$.   As
schematically illustrated in Fig.~\ref{fig:age_interpolation}, the age
interpolation tests involve combining models $n-n_{\mathrm{incr}}$ and
$n+n_{\mathrm{incr}}$, and seeing how well the interpolated frequencies reproduce
the frequencies of model $n$.  This test is carried out throughout the entire
track except for the $n_{\mathrm{incr}}$ models at either end.
Figure \ref{fig:age_interpolation} schematically illustrates these interpolation
tests for $n_{\mathrm{incr}}=2$. \textsc{aims} carries out tests for
$n_{\mathrm{incr}}=1$ and $2$ in order to assess the impact of the time step on
the age interpolation.

\begin{figure}[t]
    \centering
    \includegraphics[width=0.8\textwidth]{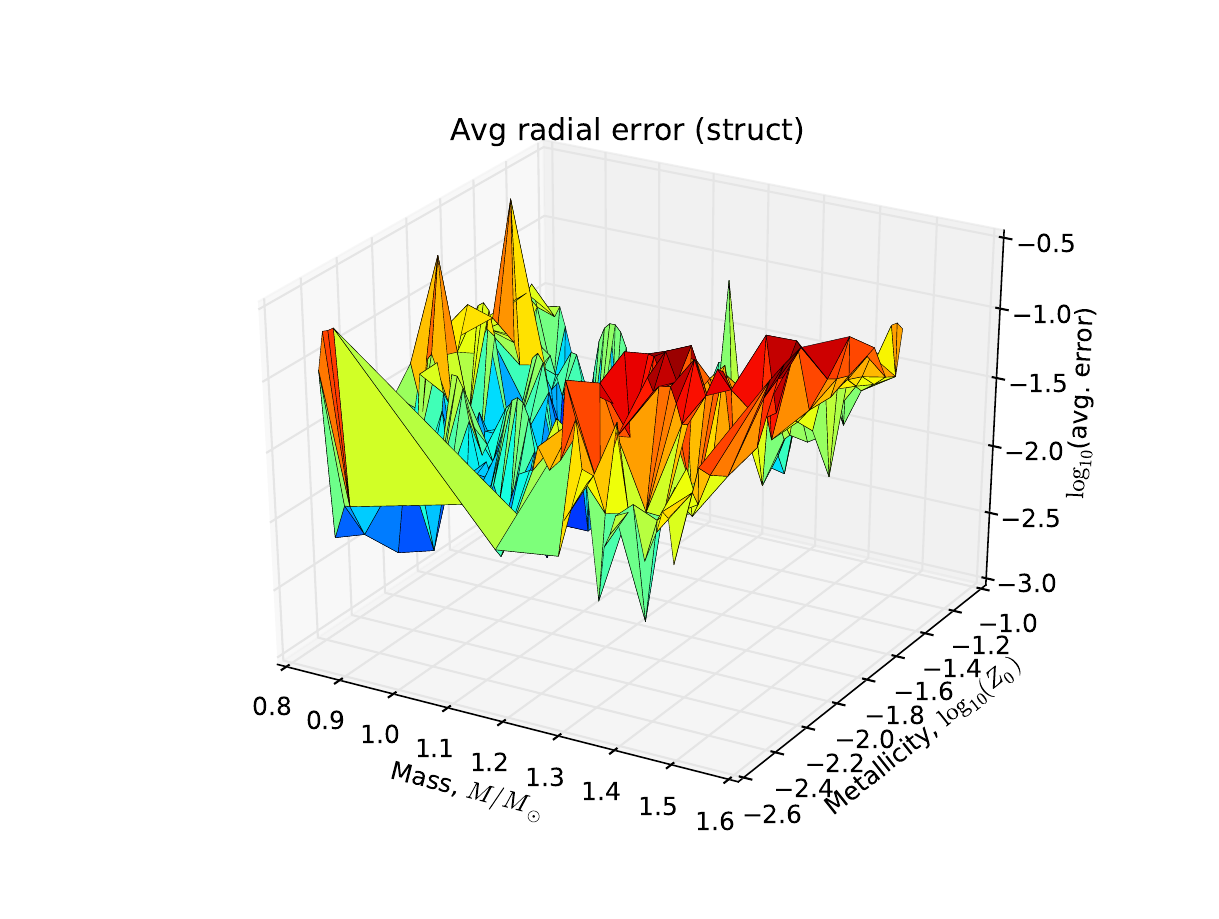}
    \caption{Average track interpolation errors for radial
    modes as a function of stellar parameters.\label{fig:3Dfigs}}
\end{figure}

Testing track interpolation (see Figs.~\ref{fig:3Dfigs} and \ref{fig:interactive2}) is more complicated because it is based on a
Delaunay tessellation.  The approach used in \textsc{aims} involves randomly
selecting half of the evolutionary tracks, creating a new tessellation from
these, and using this to interpolate to the remaining tracks.
Figure \ref{fig:interactive2} illustrates such a partitioning of the evolutionary
tracks.

When comparing frequencies from an interpolated model with those from the
original model, \textsc{aims} calculates different types of error bars.  First
of all, separate error bars are obtained for radial ($l=0$) and
for non-radial modes.  These are further subdivided into the following
categories:
\begin{itemize}
\item the maximum error;
\item a root-mean square (RMS) error;
\item an RMS error only based on the modes between $0.8\nu_{\mathrm{max}}$ and
      $1.2\nu_{\mathrm{max}}$, where $\nu_{\mathrm{max}}$ is the frequency at
      maximum power (this is obtained from the models via a scaling relation).
\end{itemize}

\subsection{Analysing the results}

The results in the generated \texttt{interpolation$\_$file} can be visualised
with\\ \texttt{plot$\_$interpolation$\_$test.py} by running:    
\begin{lstlisting}
plot_interpolation_test.py interpolation_file
\end{lstlisting}
Currently, this program only works for 3D model grids (including the age dimension).
   
Running this program will generate a series of plots that can be used to assess
the errors introduced by the interpolation.  The first 9 plots are 3D plots
which show various errors as a function of the grid parameters, excluding age. 
These plots come in groups of three: the first two plots in a group show age
interpolation errors for $n_{\mathrm{incr}}=1$ and $2$, and the third shows track
interpolation errors.  The groups of plots correspond to the following:
\begin{itemize}
\item \textbf{Plots 1--3:} Maximum interpolation errors for radial modes as a
      function of stellar parameters. For each model along a track, the maximum
      error is obtained. Then the maximum along the entire track is calculated.
\item \textbf{Plots 4--6:} Average interpolation errors for radial modes as a
      function of stellar parameters. The RMS average is calculated along the
      entire track. Figure \ref{fig:3Dfigs} shows such a plot for track interpolation
      errors.
\item \textbf{Plots 7--9:} Average interpolation errors for radial modes
      restricted to the interval $0.8\nu_{\mathrm{max}}$ to
      $1.2\nu_{\mathrm{max}}$ as a function of stellar parameters. The RMS
      average is calculated along the entire track.
\end{itemize}
These plots are displayed in individual windows.  Thanks to {\ttfamily Python}'s interactive
capabilities, it is possible to rotate the plots and to zoom in or out.

\begin{figure*}[t]
    \centering
    \begin{subfigure}
        \centering
        \includegraphics[width=0.48\textwidth]{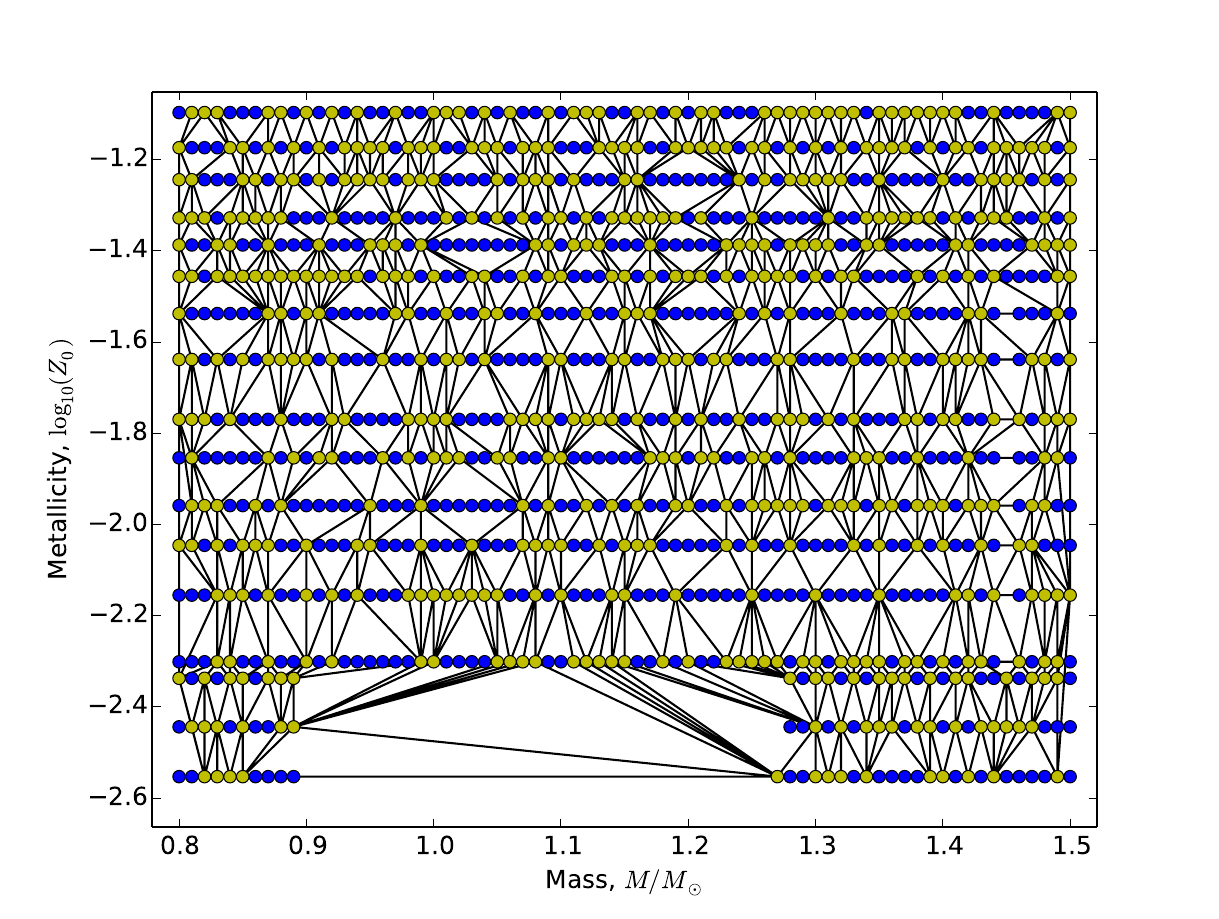}
    \end{subfigure}\hfill
    \begin{subfigure}
        \centering
        \includegraphics[width=0.48\textwidth]{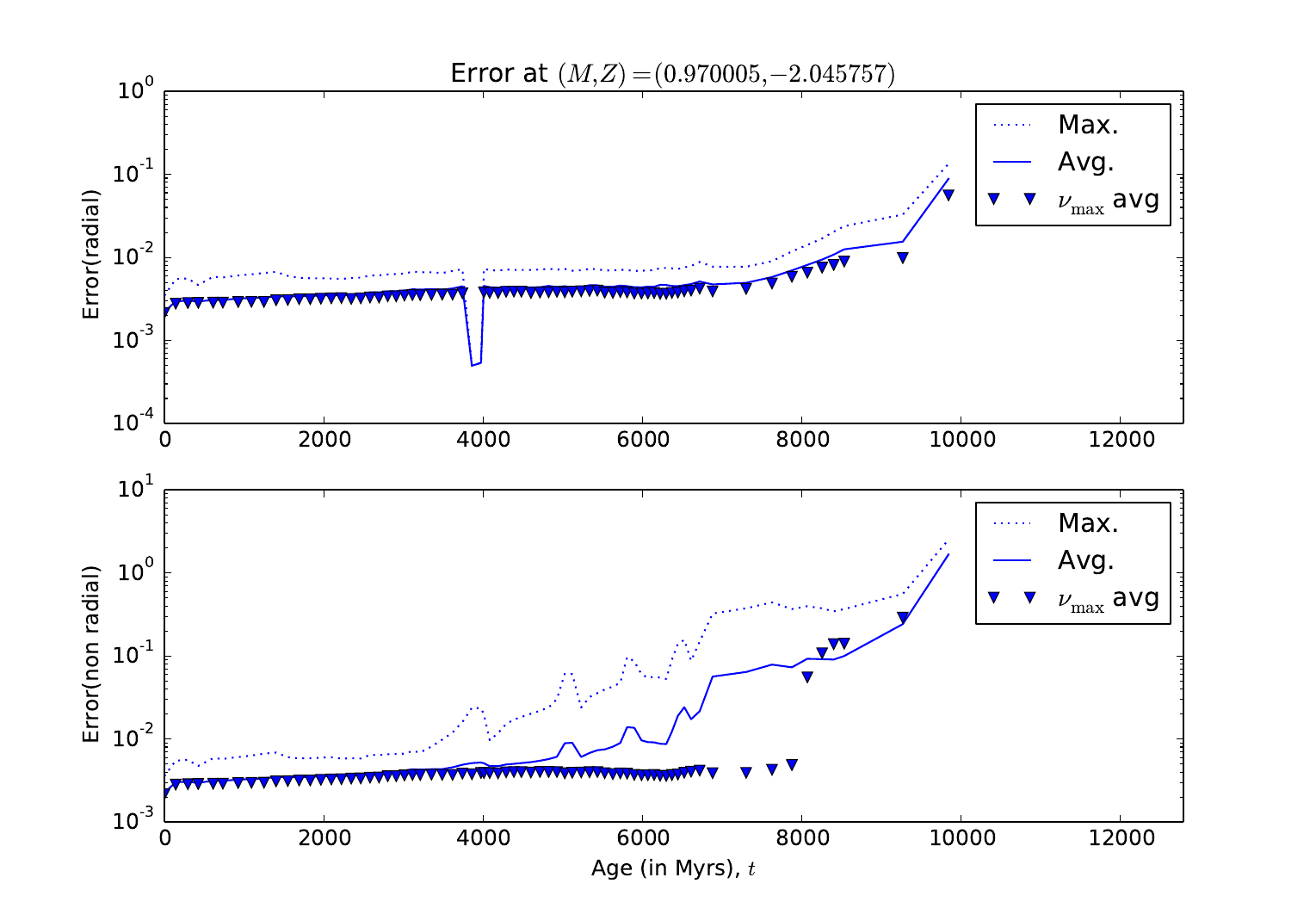}
    \end{subfigure}
    \caption{Testing interpolation accuracy. \textit{Left:} Interactive plot with the
    positions of the evolutionary tracks in parameter space.  The yellow dots
    correspond to the tracks with which a new tessellation is created, as
    represented by the connecting lines.  The blue dots represent tracks where
    the interpolation is tested. \textit{Right:} Clicking on a blue dot produces plots with the
    track interpolation errors as a function of stellar age.}
\label{fig:interactive2}
\end{figure*}

Two additional 2D interactive plots display the positions of the evolutionary
tracks in the stellar parameter space.  The first of these shows all of the
evolutionary tracks as blue dots.  Clicking on a blue dot opens up a new window
with two new plots which show how the age interpolation errors, for both radial
and non-radial modes, vary as a function of stellar age. The second plot shows a
partitioning of the evolutionary tracks used in the track interpolation tests
as described above.  An example of such a plot is shown in the left panel of
Fig.~\ref{fig:interactive2}. Clicking on a blue dot on this plot opens up a new window with two
new plots with track interpolation errors as a function of age, like the ones
shown in the right panel of Fig.~\ref{fig:interactive2}. 

More information on the individual plotting functions can be found in the
comments within the \texttt{plot$\_$interpolation$\_$test.py} file as well as at
the following website:
\url{http://bison.ph.bham.ac.uk/spaceinn/aims/version1.2/plot_interpolation_test.html}.

\section{Recommended reading}

For more information on the use of Bayesian inference in model optimisation, we
recommend \citet{2012MNRAS.427.1847B} and \citet{2013MNRAS.435..242G}. For details
on the affine-invariant MCMC optimisation scheme used (\texttt{emcee}), we refer
the reader to \citet{2010GodmanWeare} and \citet{2013PASP..125..306F}. For details on
asteroseismic grid-based analysis in general, we refer to
\citet[e.g.,][]{2011ApJ...730...63G}.

\begin{acknowledgement}
\textsc{aims} is a software for fitting stellar pulsation data, developed in the context
of the SPACEINN network, funded by the European Commission's Seventh Framework
Programme.  DRR wishes to thank all those who helped him in the development of \textsc{aims}, including D.~Bossini, T.~L.~Campante, W.~J.~Chaplin,
H.~R.~Coelho, G.~R.~Davies, B.~D.~C.~P.~Herbert, J.~S.~Kuszlewicz, M.~W.~Long,
M.~N.~Lund, and A.~Miglio.
\end{acknowledgement}

\bibliographystyle{apj}
\bibliography{biblio}

\end{document}